\documentclass[conference]{IEEEtran}
\IEEEoverridecommandlockouts
\usepackage{amsmath,amssymb,amsfonts}
\usepackage{algorithmic}
\usepackage{amsmath} 
\usepackage{graphicx}
\usepackage{textcomp}
\usepackage{xcolor}
\usepackage{amsmath}
\renewcommand{\vec}[1]{\mathbf{#1}}
\usepackage{algorithm}
\usepackage{algorithmic}
\usepackage{graphicx} 
\usepackage{float} 
\usepackage{stfloats}
\usepackage[section]{placeins}
\usepackage{subfigure}
\usepackage{booktabs}
\usepackage[numbers,sort&compress]{natbib}
\usepackage{amssymb}
\usepackage{setspace}
\usepackage{multirow}
\usepackage{color}
\usepackage{hyperref}
\usepackage{pifont}
\usepackage{bm}

\fontsize{5.0pt}{\baselineskip}\selectfont
\usepackage{epsfig}
\def\BibTeX{{\rm B\kern-.05em{\sc i\kern-.025em b}\kern-.08em
    T\kern-.1667em\lower.7ex\hbox{E}\kern-.125emX}}

\begin{document}
\addtolength{\topmargin}{0.04in}
\title{Code Optimization and Angle-Doppler Imaging for ST-CDM LFMCW MIMO Radar Systems
}
\author{Xiaolei~Shang and  Yuanbo~Cheng   
\thanks{This work was supported in part by the  National Natural Science Foundation of China under Grant 62271461.}
\thanks{X. Shang and Y. Cheng are  with the Department of Electronic Engineering and Information
Science, University of Science and Technology of China, Hefei 230027, China
(e-mail: xlshang@mail.ustc.edu.cn and cyb967@mail.ustc.edu.cn).}
}
\maketitle
\begin{abstract}
We consider code optimization and angle-Doppler imaging for  slow-time code division multiplexing  (ST-CDM) linear frequency-modulated continuous-wave (LFMCW) multiple-input multiple-output (MIMO) radar systems.  We optimize the slow-time code  via the minimization of a  Cram{\'e}r-Rao Bound (CRB)-based metric to enhance the parameter estimation performance. Then, a computationally efficient RELAX-based algorithm is presented to obtain the maximum likelihood (ML) estimates of the target angle-Doppler parameters.  Numerical examples show that the proposed approaches can be used to improve the performance of parameter estimation  and angle-Doppler imaging of ST-CDM LFMCW MIMO radar systems.  
\end{abstract}
\section{Introduction}
Millimeter-wave radar is an indispensable sensor for automotive radar  applications, including advanced driver assistance and  autonomous driving. Compared with other sensors, such as camera and lidar, radar can provide excellent sensing capabilities even in poor lighting or adverse weather conditions \cite{bilik2019rise,sun2020mimo}.  
\par  Multiple-input multiple-output (MIMO) radar  can achieve high angular resolution with a relatively  small number of antennas \cite{bliss2003multiple,li2007mimo,li2007parameter,li2009mimo}, and MIMO radar has become a standard  for  automotive applications. Because of the low cost advantage, most of the existing automotive radar systems are linear frequency-modulated continuous-wave (LFMCW) MIMO radar systems \cite{hakobyan2019high,bose2021mutual,shang2021weighted}. 

To synthesis a long virtual array, waveform orthogonality of a LFMCW MIMO radar is usually achieved by using the time division multiplexing (TDM), Doppler division multiplexing (DDM) or the slow-time code division multiplexing (ST-CDM) approaches \cite{sun2020mimo}. For both TDM and DDM, the maximum unambiguous detectable velocity will be reduced by a factor that is the number of the transmit antennas \cite{sun2020mimo}. In comparison, ST-CDM can avoid the reduction of the maximum unambiguous detectable velocity.  Moreover, ST-CDM LFMCW MIMO radar uses low-rate pseudorandom codes, changing  only  from one chirp to another. Therefore, ST-CDM  LFMCW MIMO radar is low-cost and may be perferred in diverse applications \cite{wang2020slow}. However, in the presence of Doppler shift, the ST-CDM waveform separation at each receiver cannot  be attained perfectly  via the conventional matched filtering (MF) \cite{wang2020slow,cao2018slow,bialer2021code}. The cross-correlations of the Doppler shifted slow-time waveforms can elevate the   sidelobe levels of the angle-Doppler images and drastically  reduce the radar dynamic range \cite{sun2020mimo,bialer2021code}. 

We consider herein the problem of  target angle-Doppler parameter estimation of ST-CDM  LFMCW MIMO radar. We  first derive the  Cram{\'e}r-Rao bound (CRB) to characterize its best achievable performance for an unbiased target parameter estimator. Moreover, we optimize the slow-time code  via the minimization of a CRB-based metric. Finally, a computationally efficient FFT-based  RELAX type of algorithm \cite{li1996efficient} is presented to attain the  maximum likelihood (ML) estimation of the target parameters.

\indent \textit{Notation:} We denote vectors and matrices by bold lowercase and uppercase letters, respectively. $(\cdot)^{*}$, $(\cdot)^T$ and $(\cdot)^H$ represent the complex conjugate, transpose and the conjugate transpose, respectively. $\vec{R}\in \mathbb{R}^{N\times M}$ or  $\vec{R}\in \mathbb{C}^{N\times M}$ denotes a real or complex-valued $N\times M$ matrix $\vec{R}$. $\vec{I}_N$ denotes an $N\times N$ identity matrix and $\vec{e}_n$ represents the $n$-th column of $\vec{I}_N$.
$\vec{A}\otimes\vec{B}$, $\vec{A}\odot\vec{B}$ and $\vec{A} \oplus \vec{B}$ denote, respectively,  the Kronecker, Hadamard and Khatri–Rao matrix products. ${\rm diag}(\vec{d})$ denotes a diagonal matrix with diagonal entries formed  from $\vec{d}$.  Finally, $j\triangleq \sqrt{-1}$.
\vspace{-0.1cm}
\section{Problem formulation}
\begin{figure}[htb]
    \centering
    \includegraphics[width=0.36\textwidth]{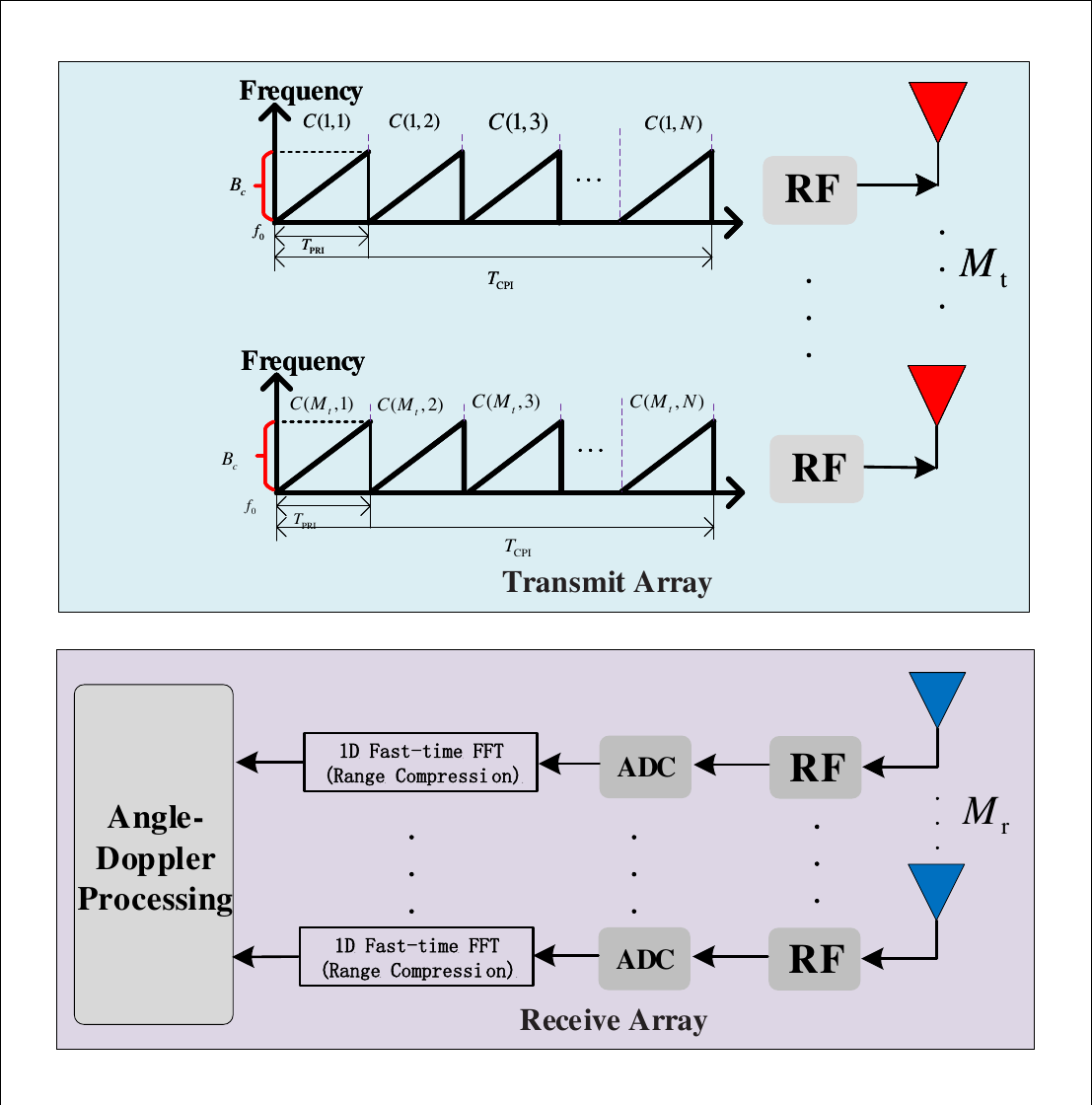}
    \caption{The ST-CDM LFMCW MIMO automotive radar architecture. $f_0$ and $B_{\rm c}$ represent the carrier frequency and bandwidth, respectively. $T_{\rm PRI}$ and $T_{\rm CPI}$  denote the duration of the chirp signal and CPI, respectively.}
    \label{fig:sysmodel}
\end{figure}

Consider a ST-CDM  LFMCW MIMO radar system with $M_{\rm t}$ transmit antennas and $M_{\rm r}$ receive antennas. As shown in Fig. \ref{fig:sysmodel}, the  MIMO radar uses $M_{\rm t}$ antennas to simultaneously transmit the same chirp after multiplying it with a unimodular random slow-time code  that is unique for each transmit antenna and  changes from one pulse repetition interval (PRI) to another within the coherent processing interval (CPI).  Let $\vec{a}_{\rm t}(\theta)$ and $\vec{a}_{\rm r}(\theta)$, respectively, denote the transmit and receive steering vectors for a target at an azimuth angle $\theta$:
\begin{align}
    \vec{a}_{\rm t}(\theta)&=\begin{bmatrix} 1, e^{-j2\pi\frac{d_{\rm t}}{\lambda}\sin\theta},  \dots, e^{-j2\pi\frac{d_{\rm t}}{\lambda}(M_{\rm t}-1)\sin\theta}\end{bmatrix}^T \in \mathbb{C}^{M_{\rm t}\times 1}, \nonumber \\
    \vec{a}_{\rm r}(\theta)&=\begin{bmatrix} 1, e^{-j2\pi\frac{d_{\rm r}}{\lambda}\sin\theta},  \dots, e^{-j2\pi\frac{d_{\rm r}}{\lambda}(M_{\rm r}-1)\sin\theta}\end{bmatrix}^T \in \mathbb{C}^{M_{\rm r }\times 1},
\end{align}
where $\lambda$ is the wavelength; $d_{\rm t}$ and $d_{\rm r}$ represent the inter-element distances of the transmit and receive uniform linear  arrays, respectively. Assume that a total number of  $N$ PRI's are used during a CPI to determine the Doppler shifts of moving targets. A nominal slow-time  temporal steering vector $\vec{d}({\omega})$ for a target with Doppler frequency shift $\omega$ can be written as:
\begin{equation}
    \vec{d}({\omega})=\begin{bmatrix} 1, e^{j\omega},  \dots, e^{j(N-1)\omega} \end{bmatrix}^T \in \mathbb{C}^{N\times 1}.
\end{equation}
As shown in Fig. \ref{fig:sysmodel}, range compression is first  performed on the received measurements.  Then, at a given fixed range bin, the data received by such a MIMO radar in the slow-time and angular domain can be written as \cite{wang2020slow,li2009mimo}:
\begin{equation}
    \vec{X}=\sum_{k=1}^K \vec{a}_{\rm r}(\theta_k)b_k (\vec{C}^T\vec{a}_{\rm t}(\theta_k) \odot \vec{d}(\omega_k))^T +\vec{E} \in \mathbb{C}^{M_{\rm r} \times N}, \label{eq:echo}
\end{equation}
where $\lbrace \theta_k \rbrace_{k=1}^K$ and $\lbrace \omega_k \rbrace_{k=1}^K$ represent the azimuth angles and Doppler shifts  of $K$ targets at a particular range bin of interest, respectively; $\lbrace b_k \rbrace_{k=1}^K$ are the  complex-valued  amplitudes, which  are proportional to the radar-cross-sections (RCS) of the targets; $\vec{C} \in \mathbb{C}^{M_{\rm t} \times N}$ represents the slow-time code matrix, with the  $(m,n)$th element $C(m,n)=e^{j\phi_{m,n}}$ denoting the slow-time code for the $m$th antenna at the $n$th PRI;  and  $\vec{E}$ is the unknown additive noise  matrix. The observed data matrix can also be expressed compactly as follows \cite{wang2020slow}: 
\begin{align}
     \vec{X}&=\vec{A}_{\rm r}\vec{B}\left( \vec{C}^T\vec{A}_{\rm t}\odot \vec{D}\right)^T + \vec{E}  \nonumber \\
    &=\vec{A}_{\rm r}\vec{B}\vec{V}^T+\vec{E}, \label{eq:model}
\end{align}
where 
\begin{align}
\vec{A}_{\rm r}&=\begin{bmatrix}\vec{a}_{\rm r}(\theta_1),\dots, \vec{a}_{\rm r}(\theta_K) \end{bmatrix} \in \mathbb{C}^{M_{\rm r} \times K}, \nonumber   \\
\vec{A}_{\rm t}&=\begin{bmatrix}\vec{a}_{\rm t}(\theta_1),\dots, \vec{a}_{\rm t}(\theta_K) \end{bmatrix} \in \mathbb{C}^{M_{\rm t} \times K}, \nonumber  \\
\vec{D}&=\begin{bmatrix}\vec{d}(\omega_1), \dots, \vec{d}(\omega_K) \end{bmatrix} \in \mathbb{C}^{N \times K}, \nonumber \\
\vec{V}&\triangleq \vec{C}^T\vec{A}_{\rm t} \odot \vec{D} \in \mathbb{C}^{N \times K}, \nonumber 
\end{align}
and
\begin{align}
\vec{b}&=\begin{bmatrix}b_1,\dots, b_K \end{bmatrix}^T, \nonumber \\
\vec{B}&={\rm diag}(\vec{b}).
\end{align}
Let the vectors $\bm{\theta}$ and $\bm{\omega}$ collect the target azimuth angles and Doppler frequencies, respectively. When there is no risk for confusion, we  omit the dependence of different functions (such as $\vec{A}_{\rm r}(\bm{\theta})$ on $\bm{\theta}$)  to simplify the notation.

Let $\bm{\varphi}$ collect all the real-valued unknown target parameters, i.e., $\bm{\varphi}=\begin{bmatrix} \bm{\theta}^T, \ \bm{\omega}^T,\ \vec{b}_{\rm R}^T, \ \vec{b}_{\rm I}^T \end{bmatrix}^T\in \mathbb{R}^{4K\times 1}$. When the noise power $\sigma^2$ is unknown,  the unknown parameter vector becomes $\bm{\zeta}=\begin{bmatrix}\bm{\varphi}^T, \ \sigma\end{bmatrix}^T\in \mathbb{R}^{(4K+1)\times 1}$. Under the assumption that $\vec{E}$ is the circularly symmetric complex-valued white Gaussian noise with  i.i.d. $\mathcal{CN}(0,\sigma^2)$ entries, the negative log-likelihood function of the measurement matrix $\vec{X}$ is given by:
\begin{equation}
    -\ln{p{(\vec{X}|\bm{\zeta})}}=\frac{ \Vert \vec{X}- \vec{A}_{\rm r}\vec{B}\vec{V}^T \Vert_{\rm F}^2}{\sigma^2} +  NM_{\rm r}\ln{\sigma^2} +NM_{\rm r}\ln{\pi}. \label{eq:7}
\end{equation}
The  Fisher information matrix (FIM) for noisy measurements $\vec{X}$ with respect to $\bm{\varphi}$ in \eqref{eq:echo} can be written as (see e.g., \cite{stoica1997}):
\begin{equation}
    \vec{F}_{0}(\bm{\varphi})=\frac{2}{\sigma^2} \Re \left\{ \left[
 \begin{matrix}
    \vec{F}_{11} & \vec{F}_{12} & \vec{F}_{13} & j\vec{F}_{13}  \\
    \vec{F}_{12}^T & \vec{F}_{22} & \vec{F}_{23} & j\vec{F}_{23} \\
    \vec{F}_{13}^T & \vec{F}_{23}^T & \vec{F}_{33} & j\vec{F}_{33} \\
   j\vec{F}_{13}^T & j\vec{F}_{23}^T & j\vec{F}_{33}^T & \vec{F}_{33} \\
  \end{matrix} 
\right] \right\}, \label{eq:F0}
\end{equation} 
where 
\begin{footnotesize}
\begin{align}
    \vec{F}_{11}&=(\dot{\vec{A}}_{\rm r}^H\dot{\vec{A}}_{\rm r})\odot (\vec{B}^{*}\vec{V}^H\vec{V}\vec{B}) +(\dot{\vec{A}}_{\rm r}^H\vec{A}_{\rm r})\odot(\vec{B}^{*}\vec{V}^H\dot{\vec{V}}_{\bm{\theta}}\vec{B}) \nonumber \\
    & \quad + ({\vec{A}_{\rm r}^H\dot{\vec{A}}_{\rm r}}) \odot (\vec{B}^*\dot{\vec{V}}_{\bm{\theta}}^H\vec{V}\vec{B})+ (\vec{A}_{\rm r}^H\vec{A}_{\rm r})\odot(\vec{B}^*\dot{\vec{V}}^H_{\bm{\theta}}\dot{\vec{V}}_{\bm{\theta}}\vec{B}), \label{eq:F11} \\
    \vec{F}_{12}&=(\dot{\vec{A}}_{\rm r}^H\vec{A}_{\rm r})\odot (\vec{B}^*\vec{V}^H\dot{\vec{V}}_{\bm{\omega}}\vec{B}) + (\vec{A}_{\rm r}^H\vec{A}_{\rm r}) \odot (\vec{B}^*\dot{\vec{V}}_{\bm{\theta}}^H\dot{\vec{V}}_{\bm{\omega}}\vec{B}), \label{eq:F12} \\
    \vec{F}_{13}&=(\dot{\vec{A}}_{\rm r}^H\vec{A}_{\rm r})\odot (\vec{B}^*\vec{V}^H\vec{V})+ (\vec{A}_{\rm r}^H\vec{A}_{\rm r}) \odot (\vec{B}^H\dot{\vec{V}}_{\bm{\theta}}\vec{V}), \\
    \vec{F}_{22}&=(\vec{A}_{\rm r}^H\vec{A}_{\rm r}) \odot (\vec{B}^*\dot{\vec{V}}_{\bm{\omega}}^H \dot{\vec{V}}_{\bm{\omega}}\vec{B}), \\
    \vec{F}_{23}&=(\vec{A}_{\rm r}^H\vec{A}_{\rm r}) \odot (\vec{B}^*\dot{\vec{V}}^H_{\bm{\omega}}\vec{V}), \\
    \vec{F}_{33}&=(\vec{A}_{\rm r}^H\vec{A}_{\rm r}) \odot (\vec{V}^H\vec{V}), \label{eq:F33}
\end{align} 
\end{footnotesize}
and
\begin{align}
    \dot{\vec{A}}_{\rm r}&=\begin{bmatrix} \frac{\partial \vec{a}_{\rm r}(\theta_1)}{\partial \theta_1}, \dots,  \frac{ \partial \vec{a}_{\rm r}(\theta_K)}{\partial \theta_K} \end{bmatrix}, \label{eq:dAr} \\ 
    \dot{\vec{V}}_{\bm{\theta}}&=\begin{bmatrix} \frac{\partial \vec{v}(\theta_1,\omega_1)}{\partial \theta_1}, \dots,  \frac{ \partial \vec{v}(\theta_K,\omega_K)}{\partial \theta_K}  
    \end{bmatrix}, \\
    \dot{\vec{V}}_{\bm{\omega}}&=\begin{bmatrix} \frac{\partial \vec{v}(\theta_1,\omega_1)}{\partial \omega_1}, \dots,  \frac{ \partial \vec{v}(\theta_K,\omega_K)}{\partial \omega_K}  
    \end{bmatrix}. \label{eq:dAf}
\end{align}
When the noise power $\sigma^2$ is unknown, the FIM for estimating $\bm{\zeta}$ is given by: 
\begin{align}
    \vec{F}_{0}(\bm{\zeta})=\left[\begin{matrix}
       \vec{F}_{0}(\bm{\varphi}) & \vec{0} \nonumber \\
       \vec{0}   &  \frac{4NM_r}{\sigma^2}
    \end{matrix}\right],
\end{align}
which is a block diagonal matrix. Hence,  the  CRB matrix for  both the   known and unknown noise power cases is given by: 
\begin{equation}
    \vec{CRB}_{0}(\bm{\varphi}) = \vec{F}_{0}^{-1}(\bm{\varphi}).
\end{equation}
In the following sections, we optimize the slow-time code via the minimization of the  CRB-based metric and present a computationally efficient RELAX-based algorithm to efficiently estimate the unknown target angle-Doppler parameters. 
\section{Slow-time Code Optimization}
The CRB matrix can be viewed as a function of the code matrix $\vec{C}$, and in practice $\bm{\varphi}$ can be replaced by the estimated target parameter vector $\hat{\bm{\varphi}}$ obtained from using an initial random slow-time code.
\par We can observe from Equations \eqref{eq:F11}$-$\eqref{eq:F33} that each block of $\vec{F}_0$ contains the products of $\vec{V}$, $\dot{\vec{V}}_{\bm{\theta}}$ and $\dot{\vec{V}}_{\bm{\omega}}$, where $\vec{V}$, $\dot{\vec{V}}_{\bm{\theta}}$,  $\dot{\vec{V}}_{\bm{\omega}}$ are all linear functions of $\vec{C}$. Therefore, $\vec{F}_0(\bm{\varphi})$ is a quadratic function of the code matrix $\vec{C}$. Making use of the property of Hadamard product $(\vec{A}\vec{C})\odot (\vec{B}\vec{D})=(\vec{A}^T\oplus \vec{B}^T)^T(\vec{C}\oplus \vec{D})$, we have 
\begin{align}
    \vec{V}^H\vec{V}&=(\vec{A}_{\rm t}^H\vec{C}^{*}\odot \vec{D}^H) (\vec{C}^T\vec{A}_{\rm t}\odot \vec{D}) \nonumber \\
    &=(\vec{A}_{\rm t}\oplus \vec{D})^H(\vec{C}^{*}\oplus \vec{I}_N)(\vec{C}\oplus \vec{I}_N)^T (\vec{A}_{\rm t}\oplus \vec{D}). \label{eq:42}
\end{align}
Similarly, the products of $\vec{V}$, $\dot{\vec{V}}_{\bm{\theta}}$ and $\dot{\vec{V}}_{\bm{\omega}}$ (i.e., $\vec{V}^H\dot{\vec{V}}_{\vec{\theta}}$, $\vec{V}^H\dot{\vec{V}}_{\bm{\omega}}$, etc.)  can also be rewritten in a similar form as in \eqref{eq:42}.  Furthermore, it is straightforward to check that:
\begin{align}
    (\vec{C}^{*}\oplus \vec{I}_N)(\vec{C}\oplus \vec{I}_N)^T&=\sum_{n=1}^N (\vec{c}^{*}_n\otimes \vec{e}_n)(\vec{c}_n^T \otimes \vec{e}_n^T) \nonumber \\
    &=\sum_{n=1}^N (\vec{c}_n^{*}\vec{c}_n^T)\otimes (\vec{e}_n\vec{e}_n^T),
\end{align}
where $\vec{c}_n$ is the $n$-th column of $\vec{C}$.  Define
\begin{equation}
    \Tilde{\vec{C}}_n=\vec{c}_n^{*}\vec{c}_n^T, \quad n=1,\dots,N,  \label{eq:C}
\end{equation}
and hence $\vec{V}^H\vec{V}$ is a linear function of the set of the rank-1 matrix $\lbrace \tilde{\vec{C}}_n\rbrace$, and so are the other terms (e.g., $\vec{V}^H\dot{\vec{V}}_{\bm{\theta}}$).  This fact implies that the FIM matrix $\vec{F}_0$ is a linear function of the rank-1 matrix $\lbrace \tilde{\vec{C}}_n\rbrace$.   It can be readily checked from Equation \eqref{eq:C} that $\tilde{C}_n(i,i)=1$ and $|\tilde{C}_n(i,j)|=1$. 
Consider next minimizing the trace of the CRB matrix:
\begin{align}
    &\min_{\tilde{\vec{C}}_n} \quad {\rm tr}(\vec{F}_0^{-1}) \nonumber \\ 
    & \ \ {\rm s.t.} \quad  \tilde{\vec{C}}_n \succeq \vec{0}, \nonumber\\
    &  \qquad \quad \  {\rm rank}(\tilde{\vec{C}}_n)=1, \quad n=1,\dots,N,  \nonumber \\
    & \qquad \quad \ \tilde{C}_n(i,i)=1, \ |\tilde{C}_n(i,j)|=1, \  i,j=1,\dots,M_{\rm t}. \label{eq:traceopt} 
\end{align}
Since \eqref{eq:traceopt} contains the non-convex  constraints ${\rm rank}(\tilde{\vec{C}}_n)=1$  and $|\tilde{C}_n(i,j)|=1$, we encounter an NP-hard optimization problem. We can relax the problem in  \eqref{eq:traceopt} by dropping the non-convex rank-one constraint and unimodular constraint:
\begin{align}
    &\min_{\tilde{\vec{C}}_n} \quad {\rm tr}(\vec{F}_0^{-1}) \nonumber \\ 
    & \ \ {\rm s.t.} \quad  \tilde{\vec{C}}_n \succeq \vec{0}, \ n=1,\dots,N,\nonumber\\
    & \qquad \quad \ \tilde{C}_n(i,i)=1, \quad n=1,\dots,N, \ i=1,\dots,M_{\rm t}. \label{eq:traceopt_relax}
\end{align}
In contrast to the original problem \eqref{eq:traceopt}, its relaxed counterpart in \eqref{eq:traceopt_relax} is a convex SDP problem and hence can be efficiently solved using interior-point methods in polynomial time \cite{helmberg1996interior,luo2010semidefinite} (e.g., using the convex optimization toolbox CVX \cite{boyd2004convex} in MATLAB).  
\par The optimal solution to the SDP problem in \eqref{eq:traceopt_relax} is denoted by $\lbrace \tilde{\vec{C}}_{n,{\rm opt}} \rbrace$, which is not necessarily rank-1 or unimodular.  We extract a rank-1 component $\vec{c}_n^{*}\vec{c}_n^T$ such that $\vec{c}_n$ is feasible for the original problem in \eqref{eq:traceopt} and serves as a good approximate solution. We perform a rank-1 approximation by taking the principal eigenvector $\bm{\upsilon}_n$ of $\tilde{\vec{C}}_{n,{\rm opt}}$ and approximate the unimodular $\vec{c}_n$ via $c_{n}(i)=\upsilon_n(i)/ |\upsilon_n(i)|$.  
\section{Maximum Likelihood Algorithm}
It follows from \eqref{eq:7} that the ML estimate of $\bm{\varphi}$ can be obtained by minimizing  the following nonlinear least squares (NLS) criterion:
\begin{align}
    \hat{\bm{\varphi}}&=\arg \min_{\bm{\varphi}} \Vert \bm{X}- \vec{A}_{\rm r}\vec{B}\vec{V}^T  \Vert _{\rm F}^2.  \label{eq:ml}
\end{align} 
The minimization of \eqref{eq:ml} is a complicated NLS problem. Herein we present a conceptually and computationally
simple RELAX-based algorithm to efficiently minimize it. Let
\begin{equation}
    \vec{X}_k=\vec{X} - \sum_{i=1,i\neq k}^{K}\hat{b}_i\vec{a}_{\rm r}(\hat{\theta}_i)\left(\vec{C}^T\vec{a}_{\rm t}(\hat{\theta}_i) \odot \vec{d}(\hat{\omega}_i) \right)^T, \label{eq:Xk}
\end{equation}
where $\lbrace \hat{b}_i, \hat{\theta}_i, \hat{\omega}_i \rbrace_{i=1,i\neq k}^K$ are assumed to be given. Then the estimate of $\lbrace b_k, \theta_k, \omega_k \rbrace$ can be obtained by minimizing the following cost function:
\begin{align}
    G(b_k,\theta_k,\omega_k)=\Vert \vec{X}_k - b_k \vec{a}_{\rm r}(\theta_k)\left( \vec{C}^T\vec{a}_{\rm t}(\theta_k)\odot \vec{d}(\omega_k)\right)^T \Vert_{\rm F}^2. 
\end{align}
A simple calculation shows that 
minimizing $G(b_k,\theta_k,\omega_k)$ with respect to  $\theta_k,\omega_k$ and $b_k$ gives:
\begin{align}
    \hat{\theta}_k,\hat{\omega}_k&= \arg \max_{\theta_k,\omega_k}   \frac{\left| \vec{a}_{\rm r}^H({\theta}_k)\vec{X}_k\vec{v}^*(\theta_k,\omega_k)\right|^2}{\Vert\vec{a}_{\rm r}(\theta_k)\Vert^2 \Vert \vec{v}(\theta_k,\omega_k)\Vert^2} \label{eq:28},  \\
    \hat{b}_k &= \frac{\vec{a}_{\rm r}^H(\hat{\theta}_k)\vec{X}_k\vec{v}^*(\hat{\theta}_k,\hat{\omega}_k)}{\Vert\vec{a}_{\rm r}(\hat{\theta}_k)\Vert^2 \Vert \vec{v}(\hat{\theta}_k,\hat{\omega}_k)\Vert^2}. \label{eq:29} 
\end{align}
It is straightforward to check that 
\begin{align}
     \vec{a}_{\rm r}^H\vec{X}_k\vec{v}^*  & = {\rm tr}\left\{ \vec{a}_{\rm r}^H\vec{X}_k\left(\vec{C}^H\vec{a}^{*}_{\rm t}\odot \vec{d}^{*} \right) \right\} \nonumber \\
    &={\rm tr}\left\{\vec{d}^{*} \odot \left(\vec{C}^H\vec{a}^{*}_{\rm t} \right) \vec{a}^H_{\rm r}\vec{X}_k \right\} \nonumber \\
    &={\rm tr}\left\{ {\rm diag}(\vec{d})\vec{C}^H\vec{a}_{\rm t}^{*}\vec{a}^H_{\rm r}\vec{X}_k  \right\}.  \label{eq:d}
\end{align}
Note that 
\begin{align}
    \vec{C}^H\vec{a}_{\rm t}^{H}\vec{a}^H_{\rm r}\vec{X}_{k} =\left[ \begin{matrix}
    \vec{c}^{H}_1\vec{a}_{\rm t}^{*}\vec{a}_{\rm r}^{H}\vec{x}_k^1 & \times & \times & \times  \\
    \times & \vec{c}^{H}_2\vec{a}_{\rm t}^{*}\vec{a}_{\rm r}^{H}\vec{x}_k^2 & \times & \times \\ 
    \vdots & \vdots & \ddots  & \vdots \\
    \times & \times &  \dots & \vec{c}^{H}_N\vec{a}_{\rm t}^{*}\vec{a}_{\rm r}^{H}\vec{x}_k^N
    \end{matrix}
    \right], \label{eq:dd}
\end{align}
where $\vec{c}_n$ and $\vec{x}_{k}^n$, respectively, represent the $n$-th columns of $\vec{C}$ and $\vec{X}_k$; and the symbol $`` \times ``$ denotes an element of no interest.   Inserting \eqref{eq:dd} into \eqref{eq:dd} yields 
\begin{align}
    \vec{a}_{\rm r}^H\vec{X}_k\vec{v}^* = \sum_{n=1}^{N} e^{-j\omega n}(\vec{a}_{\rm t} \otimes \vec{a}_{\rm r})^H(\vec{c}_n^* \otimes \vec{x}_k^n ). 
\end{align}
By uniformly gridding the angular and Doppler domains into $L_{\theta}$ and $L_{\omega}$ points,  the values of $\vec{a}_{\rm r}^H(\theta)\vec{X}_k\vec{v}^*(\theta,\omega)$
can be efficiently calculated by means of  FFT and inverse FFT operations  with zero-padding on the following matrix:
$\begin{bmatrix} \vec{c}^{*}_1\otimes \vec{x}_k^1 & \dots & \vec{c}^{*}_N\otimes \vec{x}_k^N \end{bmatrix}$. Then the corresponding angle and Doppler pair $(\hat{\theta}_k,\hat{\omega}_k)$ that maximizes the function in \eqref{eq:28} can be viewed as a coarse estimate of $\theta_k$ and $\omega_k$. We
then fine search for the solution of  \eqref{eq:28}  
over the angular interval $\left[ \hat{\theta}_k- \frac{\pi}{L_\theta}, \hat{\theta}_k + \frac{\pi}{L_\theta} \right]$ and the Doppler interval $\left[ \hat{\omega}_k- \frac{\pi}{L_\omega}, \hat{\omega}_k + \frac{\pi}{L_\omega} \right]$ by using the ‘fmincon’ function of MATLAB.  

With the above preparations, we can now present the steps of the RELAX-based algorithm for the parameter estimation of multiple targets. 

\noindent  \textit{Step 1.} Assume $K=1$. Obtain $\lbrace \hat{b}_1,\hat{\theta}_1,\hat{\omega}_1\rbrace$ from $\vec{X}$. 
\par \noindent  \textit{Step k.} Assume $K=k$. 
\begin{enumerate}
    \item Calculate $\vec{X}_k$  with \eqref{eq:Xk} via using $\lbrace \hat{b}_i,\hat{\theta}_i,\hat{\omega}_i\rbrace_{i=1}^{K-1}$ and estimate $\lbrace \hat{b}_k,\hat{\theta}_k,\hat{\omega}_k\rbrace$ by   using \eqref{eq:28} and $\eqref{eq:29}$.
    \item For $i=1:K-1$
    \begin{enumerate}
        \item[]Calculate $\vec{X}_i$  with \eqref{eq:Xk} via using $\lbrace \hat{b}_j,\hat{\theta}_j,\hat{\omega}_j\rbrace_{j=1,j\neq i}^{K}$  and reestimate $\lbrace b_i,\theta_i, \omega_i \rbrace$ from $\vec{X}_i$ by using \eqref{eq:28} and $\eqref{eq:29}$. 
    \end{enumerate}
    end 
\end{enumerate}
Iterate the previous two substeps until ‘practical convergence’ is achieved. 

\textbf{Remaining step:} RELAX stops when $K$ is equal to the estimated number of targets, which can be determined by using the Bayesian
information criterion (BIC) \cite{stoica2004model}. 

The ``practical convergence'' in Step $k$ is considered achieved  by checking the relative change of the cost function  in \eqref{eq:ml} between two consecutive iterations. In our simulations, we terminate the iterative process in Step $k$ when the aforementioned relative change is less than $10^{-3}$. 
\section{Numerical examples}
In this section, we present several numerical examples to demonstrate the performance of the proposed slow-time code optimization and parameter estimation algorithms. The LFMCW MIMO radar under consideration contains $M_{\rm t}=10$ transmit antennas spaced at $d_{\rm t}=5\lambda$ and $M_{\rm r}=10$ receive antennas spaced at $d_{\rm r}=\lambda / 2$.  The slow-time sample number $N=64$ PRI's are adopted to extract the Doppler information. In addition to our optimized codes,  the well-known constant amplitude zero auto-correlation (CAZAC) codes, such as the Zadoff-Chu sequences and P sequences, are  used as the slow-time codes for the ST-CDM LFMCW MIMO radar systems  \cite{sun2020mimo,cao2018slow}.

\begin{figure}[htb]
\centering
\subfigure[MF]{
\label{fig:MF}
\begin{minipage}[t]{0.3\linewidth}
\centering
\centerline{\epsfig{figure=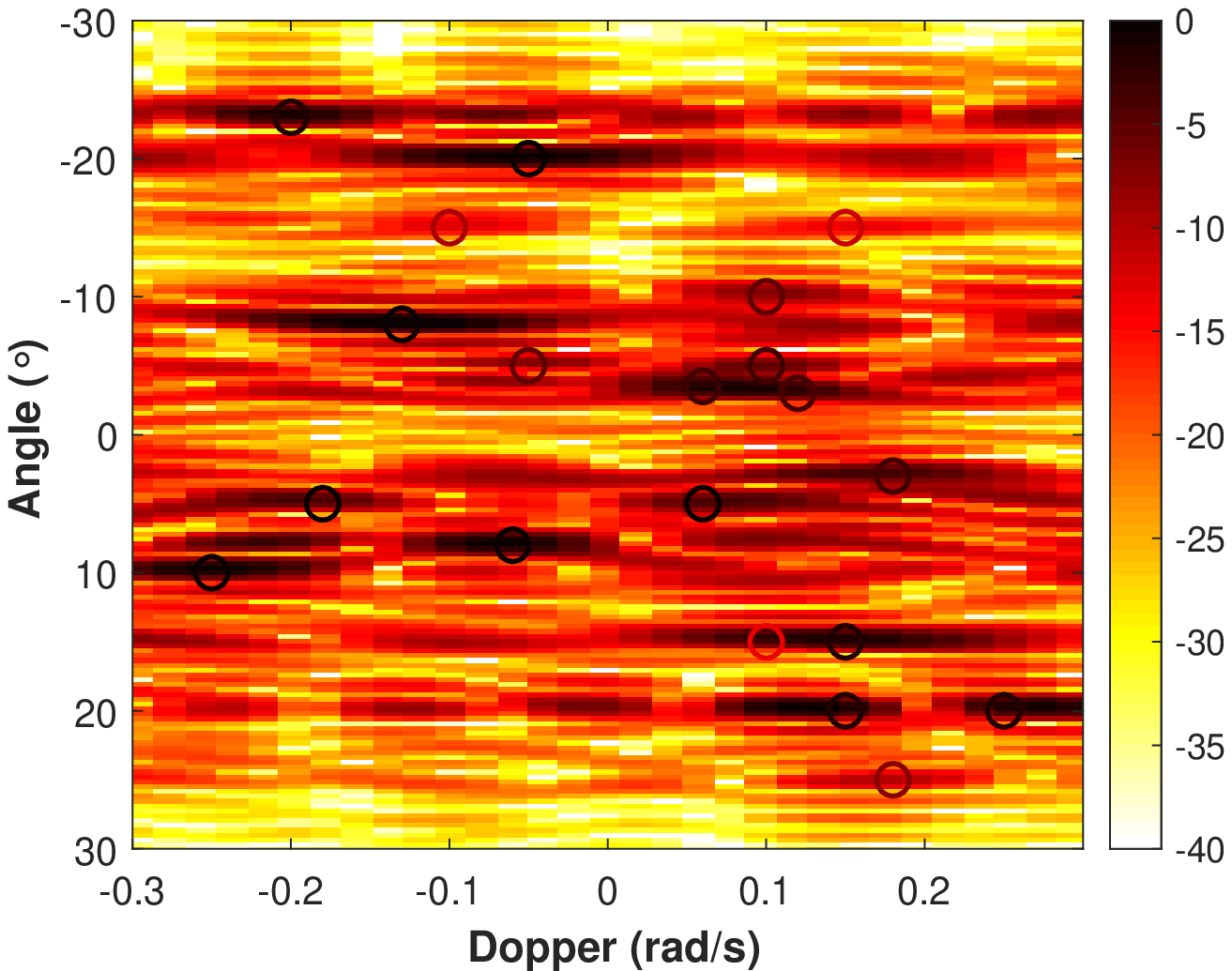,width=3.1cm}}
\end{minipage}} 
\subfigure[IAA]{
\label{fig:IAA}
\begin{minipage}[t]{0.3\linewidth}
\centering
\centerline{\epsfig{figure=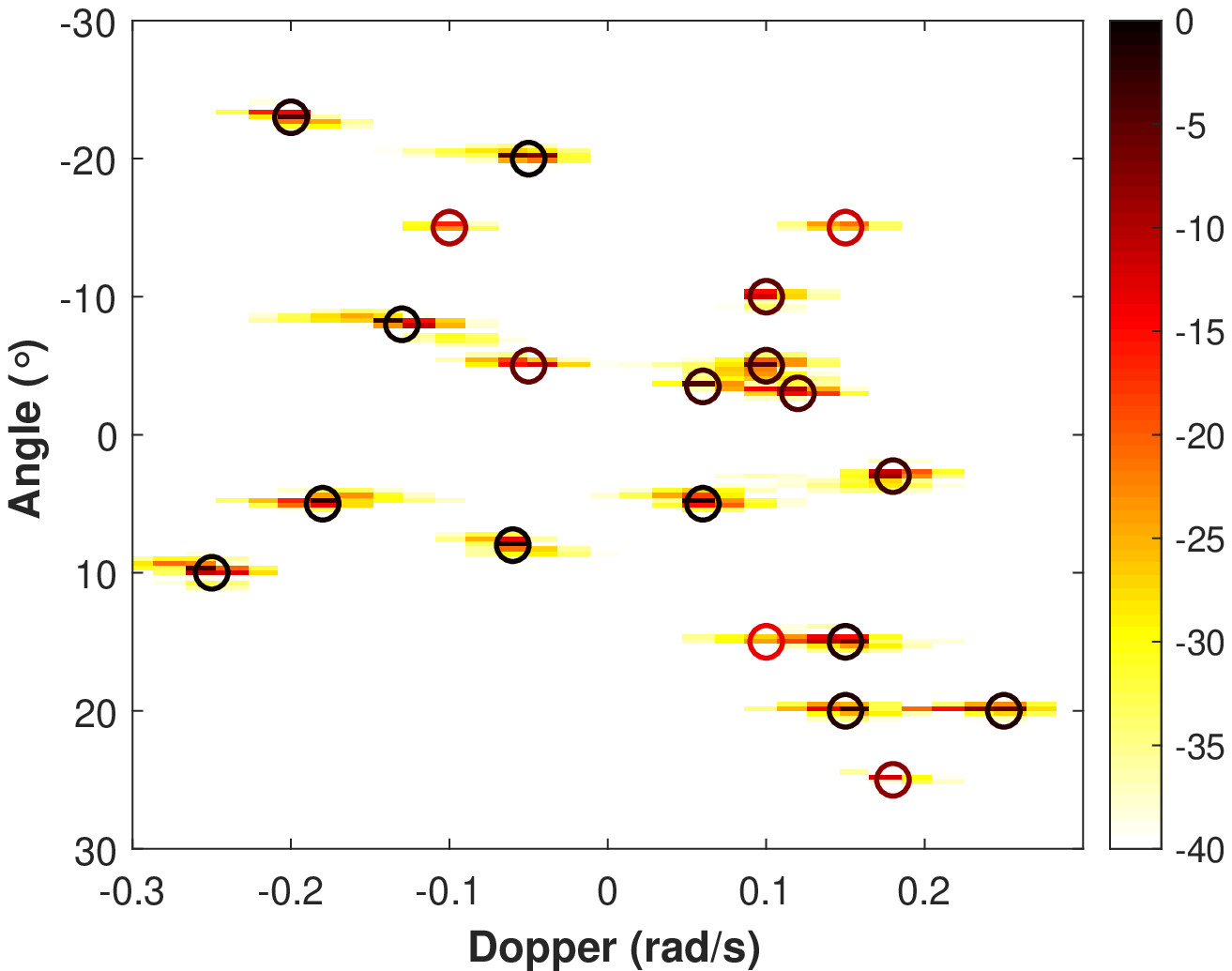,width=3.1cm}}
\end{minipage}}
\subfigure[RELAX]{
\label{fig:relax}
\begin{minipage}[t]{0.3\linewidth}
\centering
\centerline{\epsfig{figure=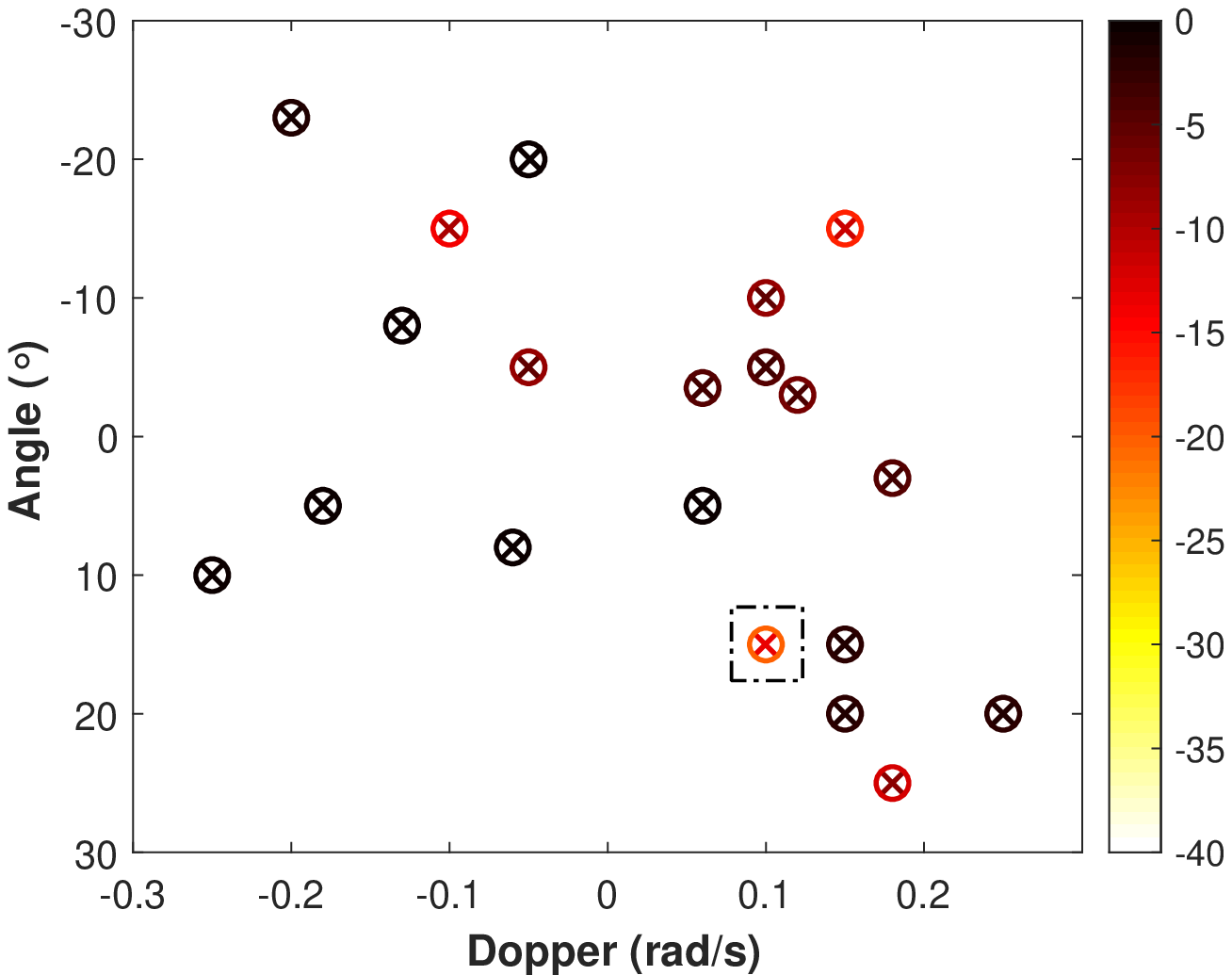,width=3.1cm}}
\end{minipage}}
\caption{Angle-Doppler images of LFMCW MIMO radar, obtained  applying (a) matched filtering, (b) IAA, and (c) the proposed RELAX-based algorithm to  the noisy measurements when Zadoff-Chu sequences are adopted as the slow-time codes.  ``$\bigcirc$'' symbols indicate  the true locations of the  targets (color-coded according to power, in dB). ``X'' symbols in Fig. \ref{fig:relax} represent targets estimated by the  RELAX-based algorithm.  } 
\label{fig:scene1}
\end{figure}

The scene of interest consists of 20 targets, with their true angle-Doppler locations and powers indicated by the color-coded `$\bigcirc$' in Fig. \ref{fig:scene1}. The SNR, which is defined as $10\log_{10}\frac{\min\lbrace|b_k|^2\rbrace_{k=1}^{20}}{\sigma^2}$, is set to 10 dB. 

Figs. \ref{fig:MF}-\ref{fig:relax}, respectively,  show the   angle-Doppler images  obtained with the matched filter (MF), iterative adaptive approach (IAA) \cite{yardibi2010source,roberts2010iterative,stoica2014weighted} and the RELAX-based algorithm when the Zadoff-Chu sequences are deployed as the slow-time codes.  From Fig.  \ref{fig:MF}, we  observe that the conventional MF results in serious smearing and high sidelobe levels problems. The super resolution IAA algorithm, as shown in Fig. \ref{fig:IAA},  provides much lower sidelobe levels than MF and identifies all targets in the scene. Finally, as shown in Fig. \ref{fig:relax},  the RELAX-based algorithm provides the most accurate estimates for all targets. 

\begin{figure}[htb]
\subfigure[Phases (in radians)]{
\label{fig:phase}
\begin{minipage}[t]{0.3\linewidth}
\centering
\centerline{\epsfig{figure=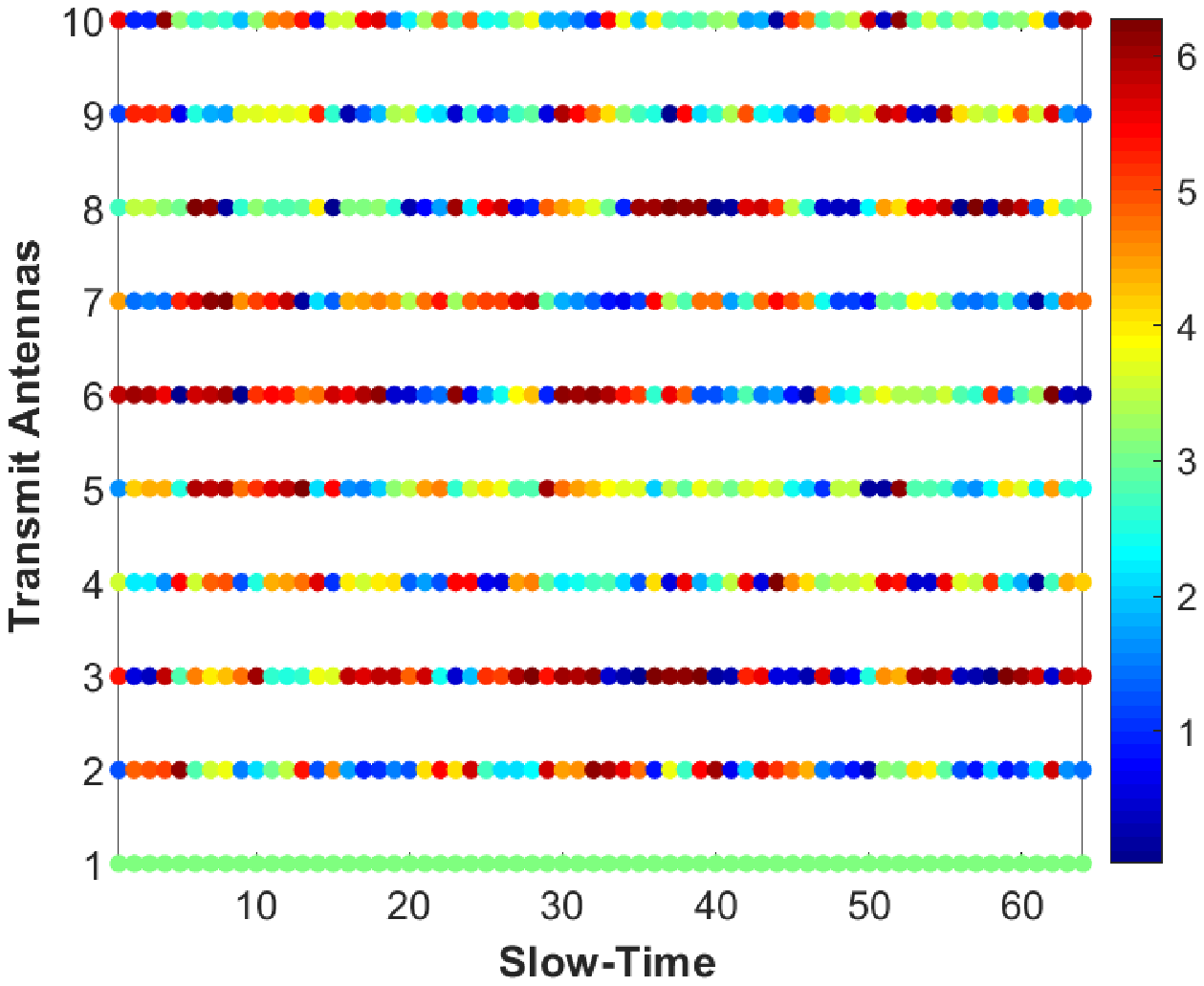,width=3.1cm}}
\end{minipage}}
\centering
\subfigure[Doppler RMSE]{
\label{fig:theta1}
\begin{minipage}[t]{0.3\linewidth}
\centering
\centerline{\epsfig{figure=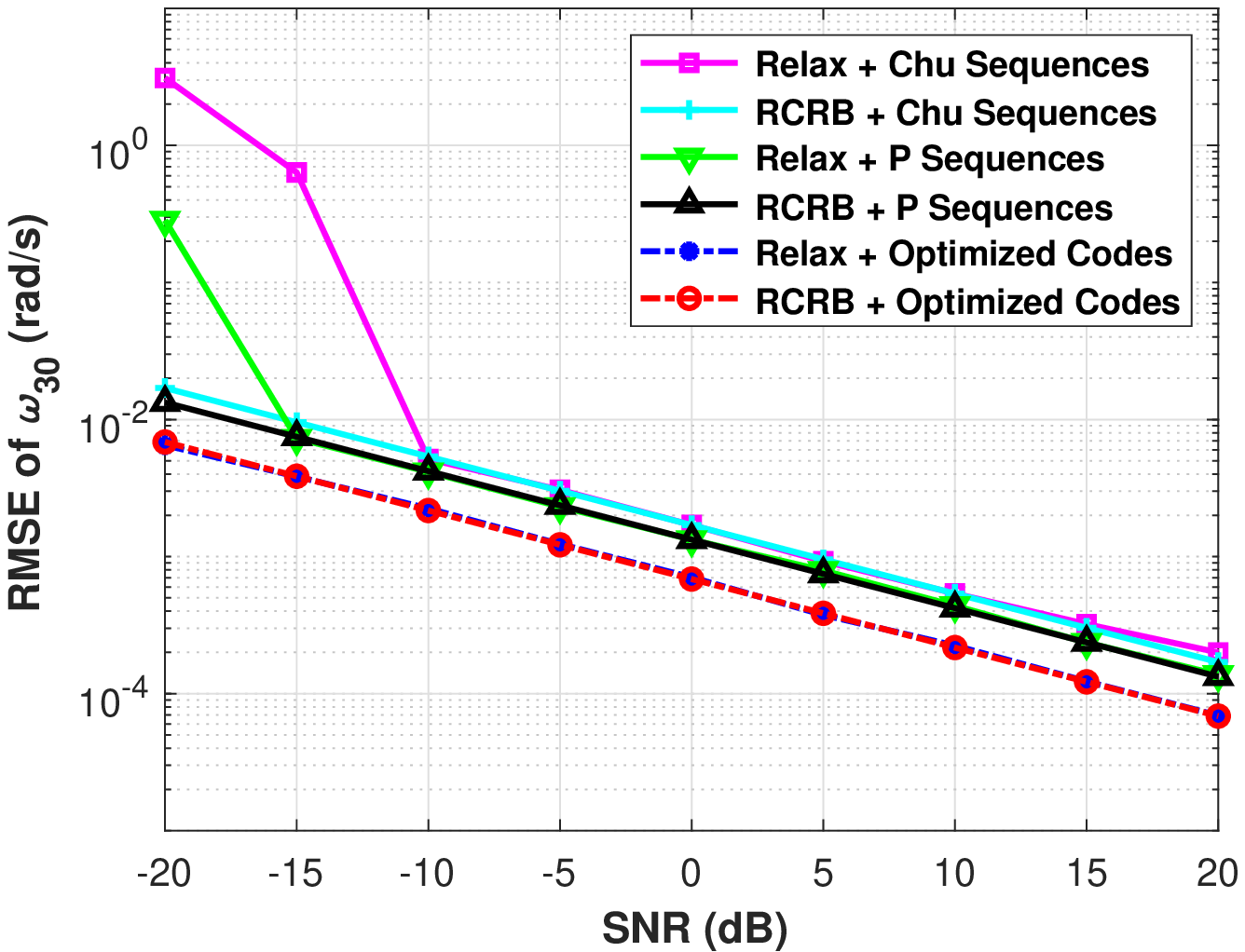,width=3.1cm}}
\end{minipage}} 
\subfigure[Angle RMSE]{
\label{fig:w1}
\begin{minipage}[t]{0.3\linewidth}
\centering
\centerline{\epsfig{figure=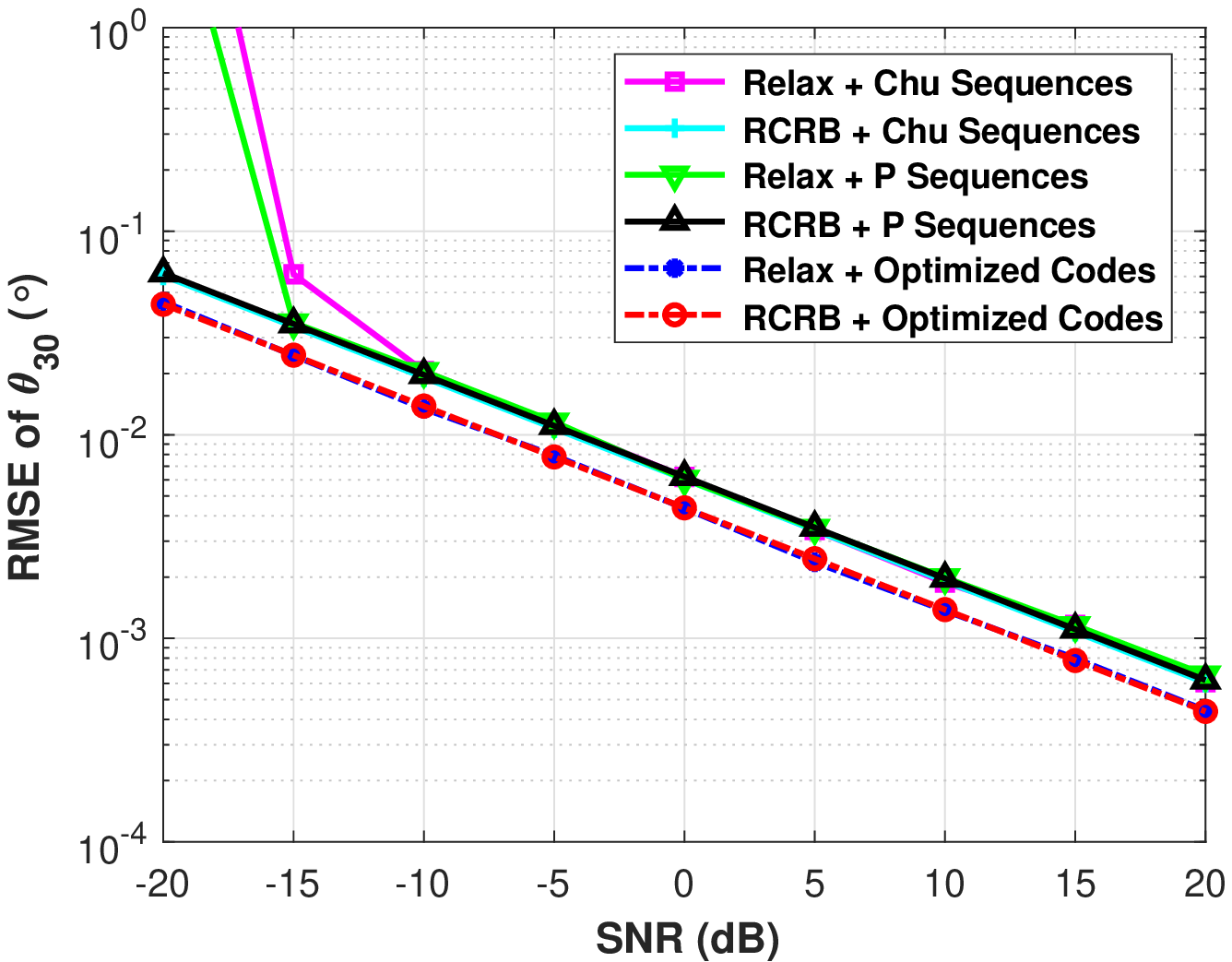,width=3.1cm}}
\end{minipage}}
\caption{RMSE comparisons among the Zadoff-Chu sequences, P sequences and the proposed optimized codes  as a function of SNR. The target of interest is marked with the dash-dot rectangle in Fig. \ref{fig:relax} with its parameters being $b_{20}=\frac{1+j}{10\sqrt{2}}, \theta_{20}=15^{\circ}, \omega_{20}=0.1 \ {\rm rad/s}$. (a) Phases of the optimized slow-time code matrix $\vec{C}$. The Doppler RMSE is shown in Fig. \ref{fig:theta1} and angle RMSE is shown in Fig. \ref{fig:w1}.   } 
\label{fig:rmse}
\end{figure}

The CRB matrix is  estimated using the initial angle-Doppler estimates in Fig. \ref{fig:relax}, and then the proposed slow-time code optimization algorithm is used
to minimize the CRB-based metric to improve the target parameter estimation performance. The phases of the optimized slow-time code matrix are ploted in Fig. \ref{fig:phase}. 

Figs. \ref{fig:theta1} and \ref{fig:w1} show the root mean-squared errors (RMSEs) of  the Doppler and angle estimates of the target of interest in Fig. \ref{fig:relax}, obtained with 500 Monte-Carlo trials and the corresponding root CRB (RCRB) as a function of SNR (see the explanations in the figure captions).   As one can see from Fig. \ref{fig:rmse}, the P sequences provide slightly better Doppler estimation performance than Zadoff-Chu sequences. Compared with Zadoff-Chu and P sequences, the optimized slow-time codes help lower the angle and Doppler RCRBs by about 2.8 dB and 7.9 dB, respectively.  
\vspace{-0.6cm}
\section{Conclusions}
We have considered the target angle-Doppler parameter estimation problem for  the ST-CDM LFMCW MIMO radar systems.  We have presented an approach to  optimize the slow-time code  via  optimizing a CRB-based metric. Then, a computationally efficient  RELAX-based algorithm was introduced to obtain the ML estimates of the unknown target angle-Doppler parameters. Numerical examples were provided to demonstrate the effectiveness of the proposed slow-time code optimization and parameter estimation algorithms.

\bstctlcite{IEEEexample:BSTcontrol}
\bibliographystyle{IEEEtran} 
\small
\bibliography{main}
\end{document}